\documentclass[aps,prl,reprint,groupedaddress]{revtex4-1}
\usepackage{graphicx}
\usepackage{color}

\newcommand{\beq}{\begin{equation}}
\newcommand{\eeq}{\end{equation}}
\newcommand{\bea}{\begin{eqnarray}}
\newcommand{\eea}{\end{eqnarray}}
\begin{document}


\title{Freeze-out parameters: lattice meets experiment}


\author{S. Borsanyi$^{1}$, Z. Fodor$^{1,2,3}$, S. D. Katz$^{2.4}$, S. Krieg$^{1,3}$, C. Ratti$^{5}$, K. K. Szabo$^{1}$}
\affiliation{$^1$ \small{\it Department of Physics, Wuppertal University, Gaussstr. 20, D-42119 Wuppertal, Germany}\\
$^2$ \small{\it Inst. for Theoretical Physics, E\"otv\"os University,}\\
\small{\it P\'azm\'any P. s\'et\'any 1/A, H-1117 Budapest, Hungary}\\
$^3$ \small{\it J\"ulich Supercomputing Centre, Forschungszentrum J\"ulich, D-52425
J\"ulich, Germany}\\
$^4$ \small{\it MTA-ELTE "Lend\"ulet" Lattice Gauge Theory Research Group,}\\
\small{\it P\'azm\'any P. s\'et\'any 1/A, H-1117 Budapest, Hungary}\\
$^5$ \small{\it Dip. di Fisica, Universit\`a di Torino and INFN, Sezione di Torino}\\
\small{\it via Giuria 1, I-10125 Torino, Italy}\\}

\date{\today}

\begin{abstract}
We present our results for ratios of higher order fluctuations of electric charge as functions of the temperature. These results are obtained in a system of 2+1 quark flavors at physical quark masses and continuum extrapolated. We compare them to preliminary data on higher order moments of the net electric charge distribution from the STAR collaboration. This allows us to determine the freeze-out temperature and chemical potential from first principles. We also show continuum-extrapolated results for ratios of higher order fluctuations of baryon number. These will allow to test the consistency of the approach, by comparing them to the corresponding experimental data (once they become available) and thus extracting the freeze-out parameters in an independent way.

\end{abstract}

\pacs{}

\maketitle

The QCD transition from a hadronic, confined system to a partonic one at zero
baryo-chemical potential is an analytic cross-over, as was unambiguously shown
by lattice QCD simulations \cite{Aoki:2006we}. This feature extends to small
chemical potentials covered by the high energy runs at RHIC. The possibility
that the transition becomes first order at large chemical potentials has
triggered the low energy runs at RHIC, soon to be followed by the CBM
experiment at the GSI, in search for the elusive critical point.  In order to
successfully spot its position, one needs to define observables which are
sensitive to the change in the order of the phase transition. Event-by-event
higher order fluctuations of conserved charges are expected to diverge in the
presence of a first order phase transition, and have therefore been proposed
long ago to this purpose \cite{Stephanov:1999zu,Gavai:2008zr,Cheng:2007jq}. As
a consequence, experimental results for these observables are becoming
available at several collision energies, covering different regions of the QCD
phase diagram \cite{McDonald:2012ts,Sahoo:2012bs}.  Recently, further interest
towards fluctuations of conserved charges and their ratios has been stimulated
even at $\mu=0$, following the idea that the freeze-out parameters can be
extracted by comparing their experimental value to lattice QCD results
\cite{Karsch:2012wm,Bazavov:2012vg}. This comparison allows to extract the
temperature and baryon-chemical potential at freeze-out from first principles,
without the need of relying on a phenomenological model such as the Hadron
Resonance Gas (HRG). This also allows to test the assumption that the
equilibrium system simulated on the lattice is suitable to describe the
experimentally measured fluctuations, since in principle non-equilibrium
effects and final-state interactions in the hadronic phase might become
relevant.  The present level of precision reached by lattice QCD simulations,
performed at physical quark masses and continuum-extrapolated, is very timely
and allows this kind of comparison between experimental data and lattice QCD
results for the first time.

In this paper we show the first continuum-extrapolated results for higher order
fluctuations of electric charge and extract the freeze-out conditions by
comparing our results to preliminary data by the STAR collaboration at RHIC
\cite{McDonald:2012ts,Sahoo:2012bs}. This follows our previous work on
second-order fluctuations of conserved charges \cite{Borsanyi:2011sw}. We also
present results for baryon number fluctuations, which can be compared to the
experimental data, once they become available (so far, only proton fluctuations
have been measured in experiments \cite{Aggarwal:2010wy}, and the issue whether
one can extract baryon number fluctuations from them is still open
\cite{Kitazawa:2012at,Bzdak:2012an}).  Our simulations are performed in a
system of 2+1 quark flavors at the physical point, i.e. with physical
$M_K/f_K$ and $M_{\pi}/f_K$ ratios at each lattice spacing, which are realized
at the strange- over light-quark mass ratio $m_s/m_{u,d}\simeq28$. 

\begin{figure}
 \scalebox{.35}{
 \includegraphics{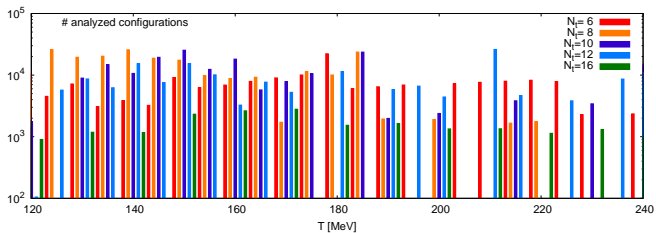}}
 \caption{Number of analyzed configurations for each temperature and each lattice spacing.
The configurations have been saved with a separation of 10 trajectories.
Each configuration was analyzed by $(128 \dots 256)\times 4$ random sources. \label{fig1}}
 \end{figure}

The continuum extrapolation is mainly performed on the basis of five lattice
spacings, corresponding to temporal lattice extents of
$N_t=6,~8,~10,~12,~16$ (around $T_c$ these extents result in lattice
spacings of $a=0.22, 0.16, 0.13, 0.11$ and $0.08$~fm, respectively).  At every lattice spacing
and temperature we analyzed every 10th configuration in the rational hybrid
Monte Carlo streams with 128\dots256 quartets of random sources.  The statistics
for each point is shown in Fig. \ref{fig1}.  We follow the extrapolation
strategy that we have discussed in Ref.~\cite{Borsanyi:2011sw}, and perform
several possible continuum fits (with and without a beyond-$a^2$ term, keeping
or dropping the coarsest lattice, using tree-level improvement
\cite{Borsanyi:2010cj} or not, fitting the observable or the reciprocal of the
observable, choosing between two possible interpolations).  Weighting these
continuum results by the goodness of the fit a histogram is formed, the width
of which defines the systematic error (for details see
Ref.~\cite{Durr:2008zz}). In this paper we show the combined systematic and
statistical errors on the continuum data.

Similarly to previous works, we choose a tree-level Symanzik 
improved gauge, and a stout-improved staggered fermionic action (see Ref.~\cite{Aoki:2005vt} for 
details). The stout-smearing \cite{Morningstar:2003gk} reduces taste violation
(this kind of smearing has one of the smallest taste violations among the ones
used in the literature for large scale thermodynamic simulations, together with
the HISQ action \cite{20,Bazavov:2010sb} used by the hotQCD collaboration).
This lattice artifact needs to be kept under control when studying higher order
fluctuations of electric charge, which are pion-dominated at small
temperatures, and thus particularly sensitive to this issue.

The observables under study are defined as:
\bea
\frac{\chi_{lmn}^{BSQ}}{T^{l+m+n}}=\frac{\partial^{\,l+m+n}(p/T^4)}{\partial(\mu_{B}/T)^{l}\partial(\mu_{S}/T)^{m}\partial(\mu_{Q}/T)^{n}}.
\eea
and they are related to the moments of the distributions of the corresponding conserved charges by
\bea
\mathrm{ mean:}~~M=\chi_1~~&&~~\mathrm{ variance:}~~\sigma^2=\chi_2
\nonumber
\\
\mathrm{ skewness:}~~S=\chi_3/\chi_{2}^{3/2}
~~&&~~
\mathrm{kurtosis:}~~\kappa=\chi_4/\chi_{2}^{2}\,.
\eea
With these moments we can express the volume independent ratios
\bea
~S\sigma=\chi_3/\chi_{2}
\quad&;&\quad
\kappa\sigma^2=\chi_4/\chi_{2}\nonumber\\
M/\sigma^2=\chi_1/\chi_2
\quad&;&\quad
S\sigma^3/M=\chi_3/\chi_1\,.
\label{moments}
\eea

The experimental conditions are such, that the three chemical potentials
$\mu_B,~\mu_Q$ and $\mu_S$ are not independent of each other: the finite baryon
density in the system is generated by the nucleon stopping in the collision
region, and is therefore due to light quarks only. Strangeness conservation
then implies that the strangeness density $\langle n_S\rangle=0$. Similarly,
the initial isospin asymmetry of the colliding nuclei yields a relationship
between the electric charge and baryon-number densities: 
\hbox{$\langle n_Q\rangle=Z/A\langle n_B\rangle$}.
For Au-Au and Pb-Pb
collisions, a good approximation is to assume $Z/A=0.4$.
%

\begin{figure}[b]
 \scalebox{.5}{
 \includegraphics{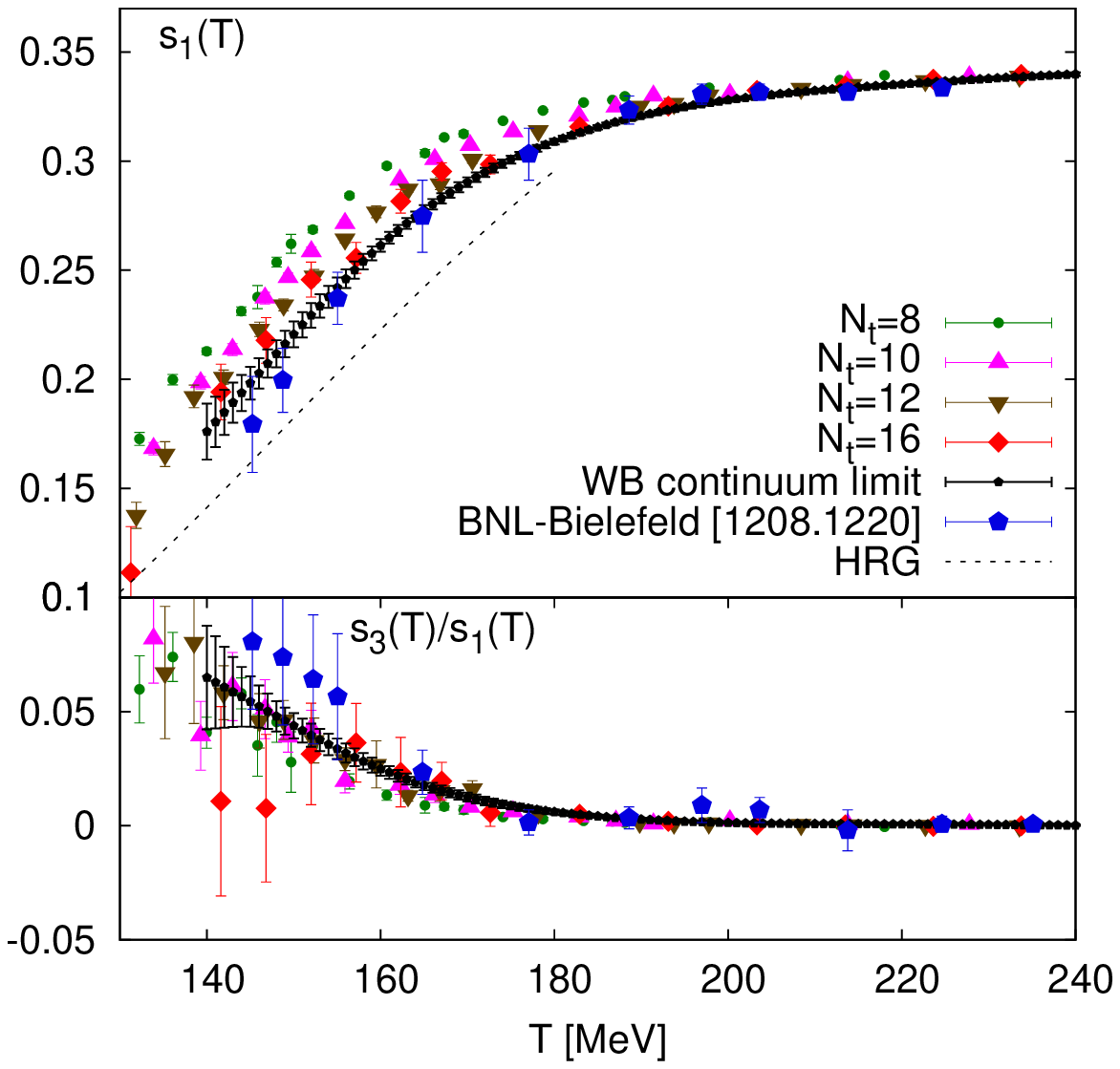}}
 \scalebox{.5}{
 \includegraphics{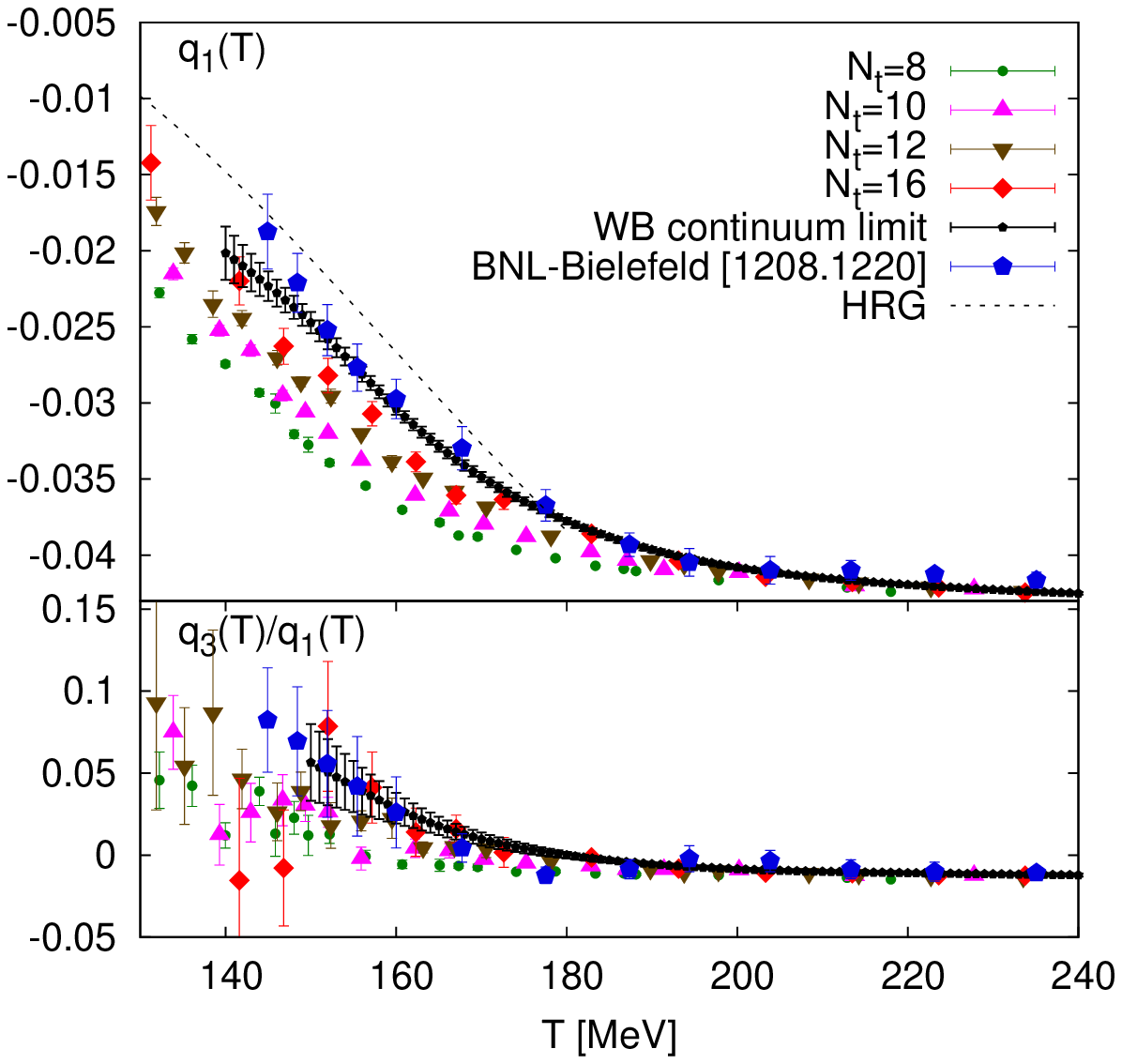}}
\caption{Upper panels: leading order contribution in $\mu_B$ for the strangeness (upper figure) and the electric charge (lower figure) chemical potentials. The lower panels show the corresponding NLO contributions. In all panels, the black dots correspond to the continuum extrapolated results. The BNL-Bielefeld results are shown as blue pentagons.}
\label{fig2bis}
\end{figure}

Therefore, the dependence of $\mu_Q$ and $\mu_S$ on $\mu_B$ needs to be defined
so that these conditions are satisfied. We take care of this by
Taylor-expanding the densities with respect to the three chemical potentials up
to order $\mu_B^3$ \cite{Bazavov:2012vg}:
\bea
\mu_Q(T,\mu_B)&=&q_1(T)\mu_B+q_3(T)\mu_B^3+...
\nonumber\\
\mu_S(T,\mu_B)&=&s_1(T)\mu_B+s_3(T)\mu_B^3+...
\label{eq:q1q2}
\eea
These equations define $q_1$, $q_3$ and $s_1$ and $s_3$, respectively.
Our continuum extrapolated data for the functions 
$q_1(T),~q_3(T),~s_1(T),~s_3(T)$ are shown
in Fig.~\ref{fig2bis}. Our data are compared to the BNL-Bielefeld group's
result, where $q_1$ and $s_1$ was continuum extrapolated. They obtained
$q_3$ and $s_3$ from $N_t=8$ lattices using the HISQ action
\cite{Bazavov:2012vg}.

The quantities that we look at, in order to extract the freeze-out temperature
and baryon chemical potential, are the ratios $\chi_3^Q/\chi_{1}^Q$ and
$\chi_1^Q/\chi_{2}^Q$ at some $(\mu_B,\mu_Q,\mu_S)$ point, which is defined
by the pyhsical conditions discussed in the previous paragraph and given by
Eq.~(\ref{eq:q1q2}). We look at ratios because they are volume-independent, and also
because they are directly related to the moments of charge distribution by Eqs.
(\ref{moments}). The first terms of their Taylor expansion around $\mu_B=0$
read:
\bea
&&R_{31}^Q(T,\mu_B)=\frac{\chi_3^Q(T,\mu_B)}{\chi_1^Q(T,\mu_B)}=
\label{LO}\\
&&\frac{\chi_{31}^{QB}(T,0)+\chi_{4}^{Q}(T,0)q_1(T)+\chi_{31}^{QS}(T,0)s_1(T)}{\chi_{11}^{QB}(T,0)+\chi_{2}^{Q}(T,0)q_1(T)+\chi_{11}^{QS}(T,0)s_1(T)}+\mathcal{O}(\mu_B^2)\nonumber
\eea
\bea
&&R_{12}^Q(T,\mu_B)=\frac{\chi_1^Q(T,\mu_B)}{\chi_2^Q(T,\mu_B)}=
\nonumber\\
&&\frac{\chi_{11}^{QB}(T,0)+\chi_{2}^{Q}(T,0)q_1(T)+\chi_{11}^{QS}(T,0)s_1(T)}{\chi_2^Q(T,0)}\frac{\mu_B}{T}+\mathcal{O}(\mu_B^3).
\nonumber
\eea
The leading order in $\chi_3^Q/\chi_1^Q$ is independent of $\mu_B$, which allows us to use $R_{31}^Q$ to extract the freeze-out temperature. Once $T_f$ has been obtained with this method, the ratio $R_{12}^Q$ can then be used to determine $\mu_B$. Notice that in Eq. (\ref{LO}) we write the expansion of $R_{12}^Q$, but in the plots we will show our results up to NLO. 
\begin{figure}
 \scalebox{.72}{
 \includegraphics{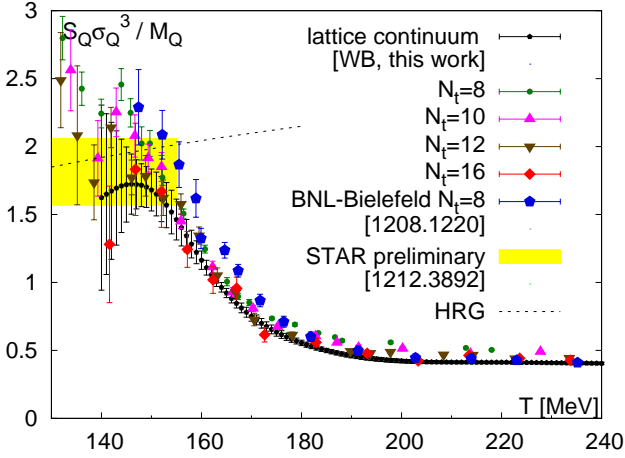}}
 \caption{
$R_{31}^Q$: the colored symbols correspond to lattice QCD simulations at finite-$N_t$.
 Black points correspond to the continuum extrapolation; blue pentagons are the $N_t=8$ results from the BNL-Bielefeld collaboration \cite{Bazavov:2012vg}. The yellow band is the preliminary STAR measurement of $S_Q\sigma_Q^3/M_Q$ \cite{Sahoo:2012bs}: it has been obtained by averaging the two most central measurements from STAR over three collision energies: $\sqrt{s}=27,~39,~62.4$ GeV. 
\label{RQ31}}
 \end{figure}
  \begin{figure}
 \scalebox{.7}{
 \includegraphics{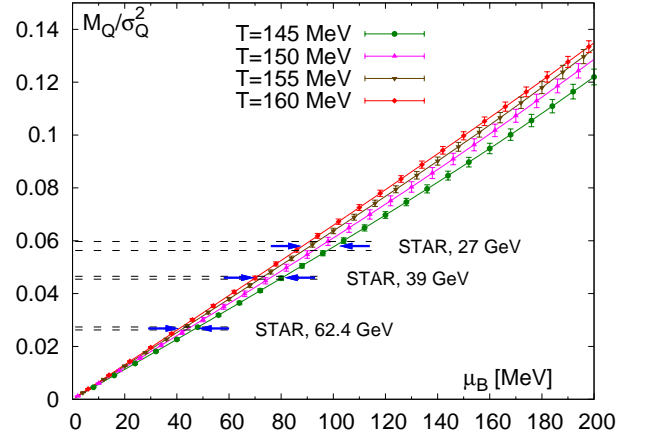}}
 \caption{$R_{12}^Q$ as a function of $\mu_B$: the different colors correspond to the continuum extrapolated lattice QCD results, calculated at different temperatures.
 The three points correspond to preliminary STAR data for $M_Q/\sigma_Q^2$ at different collision energies:   $\sqrt{s}=27,~39,~62.4$, from Ref. \cite{Sahoo:2012bs}.  \label{fig:R12Q}}
 \end{figure}

 In Fig. \ref{RQ31} we show the ratio $R_{31}^Q$ as a function of the temperature.
 The continuum extrapolation, shown in the figure as black dots, is performed
on the basis of five lattice spacings. Results from the BNL-Bielefeld
collaboration corresponding to $N_t=8$ (from Ref. \cite{Bazavov:2012vg}) are
also shown for comparison. The yellow band indicates the experimental value for
$R_{31}^Q$ from the STAR collaboration \cite{Sahoo:2012bs}. It has been
obtained by averaging the two most central measurements from STAR over three
collision energies: $\sqrt{s}=27,~39,~62.4$ GeV. We assume that this average
safely allows to determine the freeze-out temperature, since the curvature of
the phase diagram is very small around $\mu_B=0$ \cite{Endrodi:2011gv};
therefore, we expect a small variation of $T_f$ over the chemical potential
range corresponding to these three energies.  Due to the big error-bar in the
experimental measurement, and to the uncertainty in the lattice data at small
temperatures, we can only get an upper limit for the freeze-out temperature: so
far it appears that the freeze-out takes place at a temperature
$T_f\lesssim157$ MeV.
(Allowing for a two-sigma deviation both for the lattice simulation as well as the experimental data the highest possible freeze-out temperature is 161 MeV.)
 
In Fig. \ref{fig:R12Q} we show our results for $R_{12}^Q$ as a function of the
baryon chemical potential: the different curves correspond to different
temperatures, in the range of $T_f$ determined from $R_{31}^Q$.
The three STAR measurements, from Ref.  \cite{Sahoo:2012bs}, correspond to the
collision energies $\sqrt{s}=27,~39,~62.4$. Taking into account the limit on
$T_f$ that we obtained through $R_{31}^Q$, the three values of $\mu_B$ that we
extract from this observable are listed in Table \ref{tab:mu}.
The experimental evidence for the freeze-out temperature was just an
upper bound (cf. Fig.~\ref{RQ31}), thus using the data in Fig.~\ref{fig:R12Q}
can only provide for the $\mu_B$ prediction a lower bound.
In Table \ref{tab:mu} we assume that $T_f>145$ MeV. The uncertainty in the
freeze-out temperature is the dominant source of error.

 \begin{table}
  \begin{tabular}{|c|c|}
 \hline
$\sqrt{s} [GeV]$ & $\mu_B^f$ [MeV]\\
\hline
62.4&44(3)(1)(2)\\
39&75(5)(1)(2)\\
27&95(6)(1)(5)\\
&$()_{\rm \delta T}()_{\rm lat}()_{\rm exp}$\\
\hline
 \end{tabular}
 \caption{
\label{tab:mu}
Freeze-out baryon chemical potentials vs. the corresponding collision
energy of the three STAR measurements from Ref. \cite{Sahoo:2012bs}. 
The errors come from the uncertainty of the freeze-out temperature, the lattice
statistics and the experimental error, respectively.  Notice that from Fig.
\ref{RQ31} we were only able to obtain an upper limit on $T_f$.  The values of
$\mu_B$ and the error-bars in this table assume that $T_f$ is between 145
and 160 MeV, this uncertainty dominates the overall errors. (Doubling the
experimental as well as lattice errors would increase full error only by a
factor of 1.5.)
}
 \end{table}

Note that these chemical potentials differ from the results of the
statistical hadronization model \cite{Andronic:2005yp,Cleymans:2005xv}. Also the typical
freeze-out temperatures from the statistical fits lie above the upper bound
found in this work.

  \begin{figure}[b]
 \scalebox{.7}{
 \includegraphics{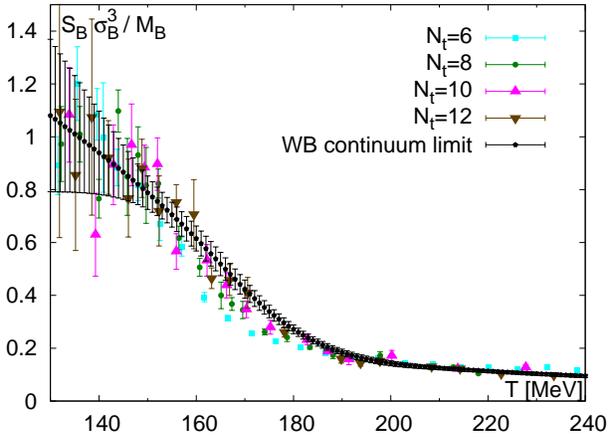}}
 \caption{$R_{31}^B$: the colored symbols correspond to lattice QCD simulations at finite-$N_t$.
 The black points correspond to the continuum extrapolation. \label{fig4}}
 \end{figure}
 
\begin{figure}
 \scalebox{.7}{
 \includegraphics{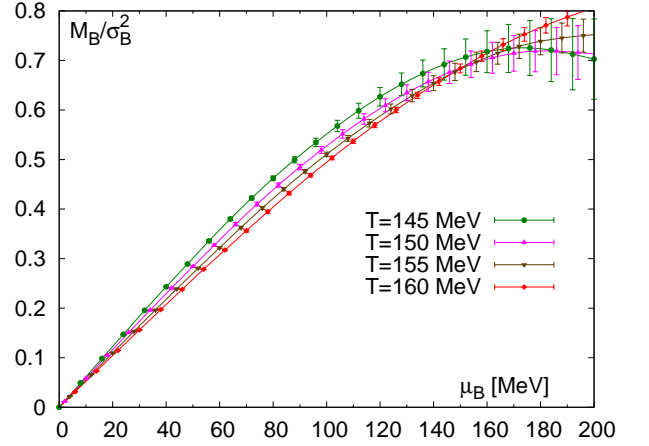}}
 \caption{\label{RB12}$R_{12}^B$: the colored symbols show
the continuum extrapolated data at various temperatures.
This quantity might also be used for $\mu_B$ measurement in the future.
}
\end{figure}

In Fig. \ref{fig4} we show our results for $R_{31}^B$ as a function of the temperature, while in Fig. \ref{RB12} we show $R_{12}^B$ for different temperatures, as a function of $\mu_B$. 
Their Taylor expansions around $\mu_B=0$ read:
\bea
&&R_{31}^B(T,\mu_B)=\frac{\chi_3^B(T,\mu_B)}{\chi_1^B(T,\mu_B)}=
\nonumber\\
&&\frac{\chi_{4}^{B}(T,0)+\chi_{31}^{BQ}(T,0)q_1(T)+\chi_{31}^{BS}(T,0)s_1(T)}{\chi_{2}^{B}(T,0)+\chi_{11}^{BQ}(T,0)q_1(T)+\chi_{11}^{BS}(T,0)s_1(T)}+\mathcal{O}(\mu_B^2)
\nonumber\\
&&
\nonumber\\
&&R_{12}^B(T,\mu_B)=\frac{\chi_1^B(T,\mu_B)}{\chi_2^B(T,\mu_B)}=
\nonumber\\
&&\frac{\chi_{2}^{B}(T,0)+\chi_{11}^{BQ}(T,0)q_1(T)+\chi_{11}^{BS}(T,0)s_1(T)}{\chi_{2}^{B}(T,0)}\frac{\mu_B}{T}
+\mathcal{O}(\mu_B^3).
\nonumber
\eea
Therefore, similarly to the electric charge fluctuations, $R_{31}^B$ allows to
extract $T_f$ and from $R_{12}^B$ we can then obtain $\mu_B$. This will allow
to independently extract the freeze-out temperature and chemical potential by
comparing them to the corresponding experimental values, once they become
available. 
Notice that the ordering of the temperatures
in Fig.~\ref{fig:R12Q} and Fig.~\ref{RB12} is opposite. $R_{12}^B$ might in
future be used to set an upper bound for $\mu_B$.
This cross-check is of fundamental importance: an inconsistency
between the two sets of freeze-out parameters obtained from the electric charge
and baryon number fluctuations might signal that it is not possible to treat
the experimental system in terms of lattice QCD simulations in thermal
equilibrium. 

\begin{figure}
 \scalebox{.72}{
 \includegraphics{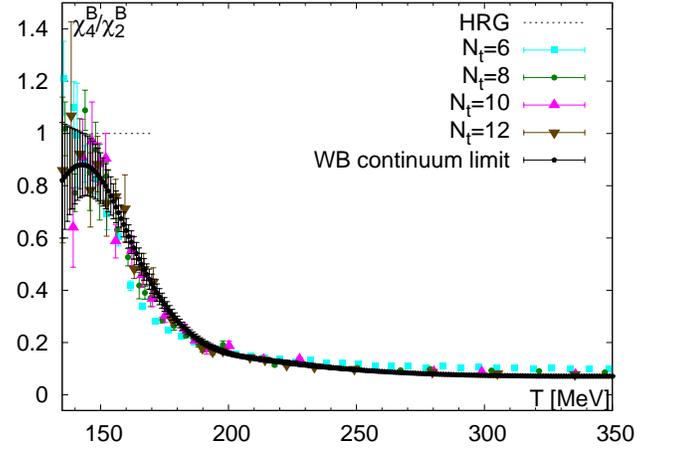}}
 \caption{$R_{42}^B$ as a function of the temperature. The black symbols correspond to the continuum extrapolation, the colored ones to the finite-$N_t$ simulations. \label{fig6}}
 \end{figure}
 In Fig. \ref{fig6} we show the ratio $R_{42}^B={\chi_4^B(T,\mu_B)}/{\chi_2^B(T,\mu_B)}$ as a function of the temperature. This observable corresponds to $\kappa\sigma^2$ of the baryon number distribution. It will allow to further independently extract $T_f$.
Notice that, in the case of baryon number, the observables are essentially flat in the hadronic phase: if the experimental value should lie in the transition region ($T\gtrsim150$ MeV) we will be able to accurately determine $T_f$,
if it lies in the hadronic phase we will only be able to provide an upper limit for the freeze-out temperature.

In conclusion, we have presented our continuum-extrapolated results for ratios
of higher-order fluctuations of electric charge and baryon number and compared
them to recently measured moments of electric charge distribution from the STAR
collaboration. This procedure has allowed us to extract, for the first time,
the values for the freeze-out parameters $T_f$ and $\mu_B^f$ from first
principles. So far it is only possible to extract an upper limit for $T_f$, due
to both experimental and lattice QCD uncertainties. The value that we obtain,
$T_f\lesssim157$ MeV, is well within the transition range predicted
from lattice QCD simulations \cite{Tctrilogy}. This is compatible with the
expectation that freeze-out occurs just below the transition \cite{BraunMunzinger:2003zz}.

\textrm{Acknowledgments:}
This project was funded by the DFG grant SFB/TR55.
The work of C. Ratti is supported by funds provided by the Italian Ministry of
Education, Universities and Research under the
Firb Research Grant RBFR0814TT.
S. D. Katz is funded by the ERC grant ((FP7/2007-2013)/ERC No 208740)
as well as the "Lend\"ulet" program of the Hungarian Academy of Sciences
((LP2012-44/2012).
The numerical simulations were performed on the QPACE machine, the GPU cluster
at the Wuppertal University and on JUQUEEN (the Blue Gene/Q system of the
Forschungszentrum Juelich).

\bibliography{biblio}
\end{document}